\documentclass[fleqn,usenatbib]{mnras}

\usepackage{newtxtext,newtxmath}
\usepackage[T1]{fontenc}
\usepackage{orcidlink}

\DeclareRobustCommand{\VAN}[3]{#2}
\let\VANthebibliography\thebibliography
\def\thebibliography{\DeclareRobustCommand{\VAN}[3]{##3}\VANthebibliography}

\usepackage{graphicx}	
\usepackage{amsmath}	
\usepackage{multirow}
\usepackage{textgreek}
\usepackage{url}
\usepackage{bm}

\defcitealias{sunyaev-zeldovich}{Sunyaev-Zel'dovich (1970)}

\newcommand\dSG{\delta_\mathrm{SG}}
\newcommand\dSI{\delta_\mathrm{SI}}
\newcommand\dDI{\delta_\mathrm{DI}}
\newcommand\rS{\bm{r}_\mathrm{S}}
\newcommand\rSG{\bm{r}_\mathrm{SG}}
\newcommand\rSD{\bm{r}_\mathrm{SD}}

\newcommand\sSG{\sigma_\mathrm{SG}}
\newcommand\sSI{\sigma_\mathrm{SI}}
\newcommand\sDI{\sigma_\mathrm{DI}}
\newcommand\sSD{\sigma_\mathrm{SD}}
\newcommand\sS{\sigma_\mathrm{S}}
\newcommand\sG{\sigma_\mathrm{G}}
\newcommand\sD{\sigma_\mathrm{D}}
\newcommand\bpar{\beta_{\parallel}}
\newcommand\bper{\beta_{\perp}}
\newcommand\soverm{\sigma/m}
\newcommand\ngal{n_\mathrm{gal}}

\title[Merging galaxy clusters and SIDM]{Hydrodynamical simulations of merging galaxy clusters:\\giant dark matter particle colliders, powered by gravity}

\author[E.\ L.\ Sirks et al.]{Ellen L.\ Sirks\,\orcidlink{0000-0002-7542-0355}$^{1,2}$\thanks{E-mail: ellen.sirks@sydney.edu.au},
David Harvey\,\orcidlink{0000-0002-6066-6707}$^{3}$,
Richard Massey\,\orcidlink{0000-0002-6085-3780}$^{2}$,
Kyle A.\ Oman\,\orcidlink{0000-0001-9857-7788}$^{2}$,
Andrew Robertson\,\orcidlink{0000-0002-0086-0524}$^{4}$,
\newauthor
Carlos Frenk\,\orcidlink{0000-0002-2338-716X}$^{2}$,
Spencer Everett\,\orcidlink{0000-0002-3745-2882}$^{4}$,
Ajay S.\ Gill\,\orcidlink{0000-0002-3937-4662}$^{5}$,
David Lagattuta\,\orcidlink{0000-0002-7633-2883}$^{2,6}$
\& Jacqueline McCleary\,\orcidlink{0000-0002-9883-7460}$^{7}$
\\
$^{1}$School of Physics, The University of Sydney and ARC Centre of Excellence for Dark Matter Particle Physics, Sydney, NSW, 2006, Australia\\
$^{2}$Institute for Computational Cosmology, Department of Physics, Durham University, South Road, Durham DH1 3LE, UK\\
$^{3}$Laboratoire d’Astrophysique, \'Ecole Polytechnique F\'ed\'erale de Lausanne (EPFL), Observatoire de Sauverny, CH-1290 Versoix, Switzerland\\
$^{4}$Jet Propulsion Laboratory, California Institute of Technology, 4800 Oak Grove Drive, Pasadena, CA 91109, USA\\
$^{5}$David A.\ Dunlap Department of Astronomy and Astrophysics, University of Toronto, 50 St.\ George Street, Toronto, ON M5S 3H4, Canada \\
$^{6}$Centre for Extragalactic Astronomy, Department of Physics, Durham University, South Road, Durham DH1 3LE, UK\\
$^{7}$Department of Physics, Northeastern University, 360 Huntington Ave, Boston, MA 02115, USA}

\date{Accepted 2024 April 10. Received 2024 April 09; in original form 2023 November 05.}

\pubyear{2024}

\begin{document}
\label{firstpage}
\pagerange{\pageref{firstpage}--\pageref{lastpage}}
\maketitle

\begin{abstract}

\noindent Terrestrial particle accelerators collide charged particles, then watch the trajectory of outgoing debris -- but they cannot manipulate dark matter. 
Fortunately, dark matter is the main component of galaxy clusters, which are continuously pulled together by gravity. 
We show that galaxy cluster mergers can be exploited as enormous, natural dark matter colliders.
We analyse hydrodynamical simulations of a universe containing self-interacting dark matter (SIDM) in which all particles interact via gravity, and dark matter particles can also scatter off each other via a massive mediator. 
During cluster collisions, SIDM spreads out and lags behind cluster member galaxies. 
Individual systems can have quirky dynamics that makes them difficult to interpret. 
Statistically, however, we find that the mean or median of dark matter's spatial offset in many collisions can be robustly modelled, and is independent of our viewing angle and halo mass even in collisions between unequal-mass systems. 
If the SIDM cross-section were $\soverm=0.1$\,cm$^2$\,g$^{-1}$\,$=$\,0.18\,barn\,GeV$^{-1}$, the `bulleticity' lag would be $\sim$5\,per~cent that of gas due to ram pressure, and could be detected at 95\,per~cent confidence in weak lensing observations of $\sim$100 well-chosen clusters.

\end{abstract}


\section{Introduction}

Galaxy clusters grow by merging with each other. During a merger, their three major constituents behave differently. Galaxies are point-like on this scale and act as collisionless test particles affected only by gravity. Diffuse gas experiences ram pressure, so is decelerated and disassociated from the galaxies. Dark matter (DM) follows a trajectory determined by whichever fundamental forces act on it. If DM interacts only via gravity, it should remain with the cluster galaxies. However, if it has a non-zero cross-section for collision with other DM particles, this self-interacting DM (SIDM) can also lag behind the galaxies \citep{clo06, Robertson_2017a}. 
Observationally, galaxies are visible in (a smoothed map of their) optical emission, while the diffuse gas is visible in X-ray emission or via the \citetalias{sunyaev-zeldovich} effect. The DM can be mapped via gravitational lensing.

The Bullet Cluster (1E0657-558) is the best known example of colliding clusters. The `Bullet' refers to the smaller cluster, which has passed through and is now moving away from the main cluster. Early weak lensing measurements of the offset between its galaxies and DM implied a self-interaction cross-section per unit mass $\soverm < 5\,\mathrm{cm^{2}\,g^{-1}}$, when calibrated against an analytic model of SIDM dynamics \citep{Markevitch_2004}. This was improved by \cite{Randall_2008} who ran SIDM-only simulations of Bullet Cluster-like systems. Combined with higher precision strong lensing measurements, \cite{Bradac_2008} found $\soverm < 1.25\,\mathrm{cm^{2}\,g^{-1}}$. Including ordinary matter in simulations of SIDM reduces its effect, by steepening the gravitational potential well at the cluster core \citep[e.g.][]{Mastromarino_2023}, and fully hydrodynamic simulations of the Bullet Cluster relaxed the constraint to $\soverm < 2\,\mathrm{cm}^{2}\,\mathrm{g}^{-1}$ \citep{Robertson_2017a}.

Particle colliders do not stop collecting data after one event. Astrophysical constraints should improve with statistical measurements from a large sample of merging clusters \citep{Massey_2011}. Furthermore, because the average velocity $v$ of DM particles increases with halo mass, measurements of collisions between galaxy clusters, galaxy groups, or individual galaxies could also characterise any velocity-dependence of the interaction: this would constrain the mass of the force mediator particle \citep{ARreview}. In a first attempt at this measurement, \citet{Harvey_2015} adopted a strategy of analysing as many cluster mergers as possible, all with equal weight. When calibrated against an analytical model, the 30 mergers in the {\it Hubble Space Telescope} ({\it HST}) archive at the time yielded a constraint on the cross-section of $\soverm<0.47\,\mathrm{cm}^{2}\,\mathrm{g}^{-1}$ at $v \sim 1000$\,km\,s$^{-1}$. In the future, this strategy can be easily extended to all-sky, monochromatic lensing surveys like {\it Euclid}. With additional telescope time, \citet{Wittman_2018} showed that multi-colour imaging can be used to reduce noise by better identifying components of clusters that entered a merger together. Some systems give anomalously high measurements; some anomalously low. Simultaneously re-weighting to account for the fact that some have more statistical power than others, the constraint changed to $\soverm<2\,\mathrm{cm}^{2}\,\mathrm{g}^{-1}$. 

To calibrate future observations, this paper uses hydrodynamical simulations of galaxy clusters with both SIDM and ordinary matter, in a cosmologically expanding volume. We study merging clusters in simulated universes with different DM interaction strengths \citep[always with a massive mediator particle;][]{Fischer_2022}, and test whether an observable offset between DM and stars could indeed be used to measure the interaction cross-section between DM particles.

\begingroup
\setlength{\tabcolsep}{2.5pt}
\renewcommand{\arraystretch}{1.2}
\begin{table*}
\centering
\caption{Properties of BAHAMAS simulated clusters that have subhaloes with mass $>5$~per~cent the total cluster mass within a sphere of radius 4\,physical~Mpc from the centre of potential. }
\label{tab:no_subs}
\begin{tabular}{lcccccccccc} 
\hline
Simulation & \multicolumn{6}{c}{Number of clusters with number of subhaloes} & \multicolumn{2}{c}{Mean mass} & \multicolumn{2}{c}{Mean separation between stars and gas} \\
~ & $N_\mathrm{clusters}$ &  $N_\mathrm{sub}=1$ &  $N_\mathrm{sub}=2$ &  $N_\mathrm{sub}=3$ &  $N_\mathrm{sub}=4$ & $N_\mathrm{sub, tot}$ & $\langle M_\mathrm{cl}\rangle$ [$10^{14}\,M_\odot$] & $\langle M_\mathrm{sub}\rangle$ [$10^{14}\,M_\odot$] & $\langle\dSG\rangle_\mathrm{cl}$ [kpc] & $\langle\dSG\rangle_\mathrm{sub}$ [kpc]
\\\hline
CDM & 107 & 82 & 20 & 4 & 1 & 138 & $2.96\pm 0.14$ & $0.39\pm 0.02$ & $22.98\pm 3.44$ & $21.70\pm 1.87$ \\
SIDM0.1 & 103 & 79 & 19 & 2 & 3 & 135 & $3.11\pm 0.21$ & $0.40\pm 0.02$ & $24.45\pm 2.79$ & $25.01\pm 2.51$ \\
SIDM0.3 & 102 & 76 & 23 & 2 & 1 & 132 & $3.15\pm 0.21$ & $0.42\pm 0.03$ & $23.43\pm 2.82$ & $25.16\pm 2.97$ \\
SIDM1 & 105 & 83 & 17 & 3 & 2 & 134 & $3.16\pm 0.20$ & $0.42\pm 0.03$ & $29.56\pm 3.83$ & $21.57\pm 2.12$ \\\hline
\end{tabular}
\end{table*}
\endgroup

The value of the SIDM cross-section is unconstrained across many orders of magnitude \citep{sidm_model_qball,sidm_model_axion,sidm_model_yukawa,maxSIDM}. A `natural' scale for models invoking a dark-sector analogue of the strong force \citep[e.g.][]{sidm_model_mirror,sidm_model_mirror2,sidm_model_hiddensector} is the same order of magnitude as nuclear interactions, $\soverm \sim 0.6\,\mathrm{cm}^{2}\,\mathrm{g}^{-1}=1$\,barn\,GeV$^{-1}$. A particularly meaningful and potentially achievable goal is to test whether $\soverm$ is significantly more or less than $0.1\,\mathrm{cm^{2}\,g^{-1}}$. If it is greater than this, DM particles that scatter off of each other are gradually ejected from dense regions of the Universe, reducing the density in the centres of haloes and slowing their gravitational collapse \citep{pet13,ethos,TYreview}. From the perspective of Occam's razor, this effect would then be sufficient to solve the seemingly unrelated `small-scale crisis' in the standard model of cosmology -- that simulations produce too much substructure that is too dense \citep{zav13,vog14,elb15}. The long time that it takes for scattering to fully erode a DM cusp would also provide a natural mechanism \citep{cre17} to explain the observed {\em diversity} of DM density profiles \citep{oman15,oldham18}. For full reviews of SIDM, see \cite{ARreview} or \cite{TYreview}.

This paper is organised as follows: we present our suite of cosmological simulations in Section~\ref{sec:sim_data}, and our methods for locating different types of matter in Section~\ref{sec:method}. We present results in Section~\ref{sec:merging_results}, including prospects for future observations. We summarise and conclude in Section~\ref{sec:conclusions}.

\section{Data}\label{sec:sim_data}
\subsection{The BAHAMAS simulations}\label{sec:bahamas}

We use the BAHAMAS suite of cosmological simulations \citep[][]{McCarthy_2017}. These use a modified version of the GADGET-3 code to model DM and baryonic physics including radiative cooling, star formation, chemical evolution, and stellar and AGN feedback.
Each simulation volume is a periodic box, 400 $h^{-1}$\,Mpc on a side. This contains $2 \times 1024^{3}$ particles, with DM particles of mass $m_{\mathrm{DM}}=5.5\times10^{9}\,\mathrm{M}_{\odot}$, gas particles initially of mass $1.1\times10^{9}\,\mathrm{M}_{\odot}$, and gravitational softening length $h_\mathrm{grav}$ that is fixed in comoving coordinates at $z>3$ then constant at $h_\mathrm{grav}=5.7$~physical kpc thereafter. They assume the WMAP 9-year cosmology \citep[$\Omega_\mathrm{m} = 0.2793$, $\Omega_\mathrm{b} = 0.0463$, $\Omega_\mathrm{\Lambda} = 0.7207$, $\sigma_{8} = 0.812$, $n_{s} = 0.972$ and $h = 0.700$;][]{Hinshaw_2013}.

SIDM with velocity-independent interaction cross sections per unit mass of $\sigma/m=[0, 0.1, 0.3, 1]\,\mathrm{cm}^{2}\,\mathrm{g}^{-1}$ is implemented in resimulations from identical initial conditions \citep[hereafter `CDM', `SIDM0.1', `SIDM0.3' and `SIDM1' runs;][]{Robertson_2019}. These values span the range of empirically-allowed cross-sections. SIDM particle scattering is infrequent, elastic, isotropic, and happens during each simulation time-step $\Delta t$, with neighbours inside radius $h_\mathrm{SIDM}=h_\mathrm{grav}$ with probability
\begin{equation}\label{eq:scatter_prob}
P_\mathrm{scat} = \frac{(\sigma/m) ~ m_\mathrm{DM} \, v \, \Delta t}{\frac{4}{3}\pi h_\mathrm{SIDM}^{3}}~,
\end{equation}
where $v$ is the particles' relative velocity. For more details about our implementation of scattering, see \cite{Robertson_2017b}.

\subsection{Colliding cluster sample selection}\label{sec:merging_sample}

In each simulation volume we identify the 300 most massive clusters in the simulation snapshot at redshift $z=0$, then select those with one or more subhaloes of $\geqslant5$~per~cent the total mass $M_\mathrm{cl}$ within 4\,Mpc. This yields ${\sim}$100 clusters and ${\sim}$135 subhaloes in each simulation (Table~\ref{tab:no_subs}). To investigate the effects of DM self-interactions on both scales, we shall analyse both the main cluster haloes and subhaloes. 

A similarly inclusive selection strategy could be employed by a future analysis of all-sky surveys. Without having selected simulated systems based on their dynamics, we find relatively small separations between their components of matter. Denoting the 3D distance between galaxies (`stars') and gas as $\dSG$, our sample has mean component separation $\langle\dSG\rangle=33\pm2$\,kpc, with rms scatter 49\,kpc (which will later be needed as variance $\langle\dSG^2\rangle=(2.4\pm 0.3)\times 10^{3}$\,kpc$^{2}$). We also find that more subhaloes have yet to reach first pericentre within their host cluster than have passed it. 

Previous studies with finite telescope time have instead preferentially observed systems with high $\dSG$ (because these turn out to be most sensitive to DM interactions: see Section~\ref{sec:averaging}). For example, \cite{Harvey_2015} selected clusters with bimodal distributions of X-ray emission {\it and} the largest separations between galaxies and gas that fitted within the field of view of the Hubble Space Telescope (HST). Their observed sample had 2D separations with mean $\langle\dSG\rangle=83\pm9$\,kpc with rms 114\,kpc or variance $\langle\dSG^2\rangle=(12.9\pm 3.0)\times10^3$~kpc$^2$ (alternatively, at mean redshift $\langle z\rangle=0.4$, $\langle\dSG\rangle=17.7\pm2.1$~arcsec with rms 25~arcsec or variance $\langle\dSG^2\rangle=632\pm157$~arcsec$^2$). 
Future studies adopting a similar selection strategy should easily be able to maintain or increase these values, because new clusters with large separations between galaxies and gas continue to be found in X-ray or SZ surveys \citep{Kubo_2009,okabe_10,tempel_17,haines_18,zeneto_20,Fu2024}. Notably, this includes nearby clusters whose separations appear huge on the sky but whose gravitational lensing signal is spread over an area too large to be observed easily by HST \citep[e.g.][]{McCleary2020}. 
Nearby clusters are particularly promising for our test. Unlike weak lensing measurements of DM \textit{mass} (where signal-to-noise follows lensing sensitivity in peaking at redshift $z$$\sim$$0.3$), weak lensing measurements of DM \textit{position} (in kpc) are optimal at $z$$\sim$$0.05$ so long as the telescope has a sufficiently large field of view to capture the broader (in arcminutes) shear field \citep{Kubo_2007,Massey_2011}.

\section{Method}\label{sec:method}
It is possible to quantify the offset between a cluster's different components using their position \citep{Massey_2011} or quadrupole moments \citep{McDonald_2022}. We choose the former, and first need a definition of position. In observational studies, the methods to find the positions of the gas, galaxies and DM all differ. In simulations, we can split the particles by type and access their distributions directly.

\subsection{Measuring the location of components of matter}\label{sec:shrinking_spheres}

We use the \textit{shrinking-spheres} method to determine the 3D location of each (star, gas, DM) component \citep[see e.g.][]{Power_2003}. A first sphere is constructed at the centre of potential returned by \textsc{subfind} \citep{Springel_2001,Dolag_2009}, with initial radius $0.35R_{200}$ for each halo or its equivalent for each subhalo\footnote{We use an initial radius for subhaloes $R_\mathrm{init} = 0.35(M_\mathrm{sub}/\frac{4}{3}\pi\,\Delta_\mathrm{c}\, \rho_\mathrm{crit})^{1/3}$, where $M_\mathrm{sub}$ is the mass of the subhalo as determined by \textsc{subfind}, $\rho_\mathrm{crit}$ is the critical density, and the overdensity constant $\Delta_\mathrm{c}=200$.}. The centre is then moved to the centre of mass of particles of a given type within the current sphere, and the radius is shrunk by a factor $f=0.9$. This process is repeated until the sphere would contain fewer than 100 particles of that type. The recorded position is the centre of mass of all particles of that type within the final sphere.

The {\it shrinking-spheres} method occasionally fails, by meandering to an incorrect local peak. Failures happen most frequently for gas particles, and more often in main haloes than subhaloes. The effect creates what appears to be either a mismatch between stellar and gas clumps that were not together at the start of infall, or the misidentificiation of a centroid analogous to that found by \cite{George_2012} for real observations of stellar light. 
Such failures lead to a 3D separation between stars and gas much larger than the typical values (of order 50\,kpc), and we mitigate them by excluding from all further analysis the $\sim$1 halo per simulation box for which we measure $\dSG>250$\,kpc. The precise value of this cut is fairly arbitrary and does not affect our results.

In observational studies, only 2D projected positions can be measured. We project the measured 3D positions along $x$, $y$ and $z$-axes by discarding one coordinate in turn, then record three independent configurations for each system. If we instead use a {\it shrinking-circles} measurement as specified by \cite{Robertson_2017a}, we find results with consistent mean and uncertainty, but which move around within the full extent of 68~per~cent error bars. This suggests that the two measurements are effectively independent, with noise that is dominated by chance projection of substructures. 
We shall carry out analyses in both 3D and different implementations of 2D, but no such choices affect our final conclusions.

\subsection{Measuring spatial offsets between components of matter}\label{sec:offsets}

\begin{figure}
\centering
\includegraphics[trim={0 10mm 0 2mm},width=\linewidth]{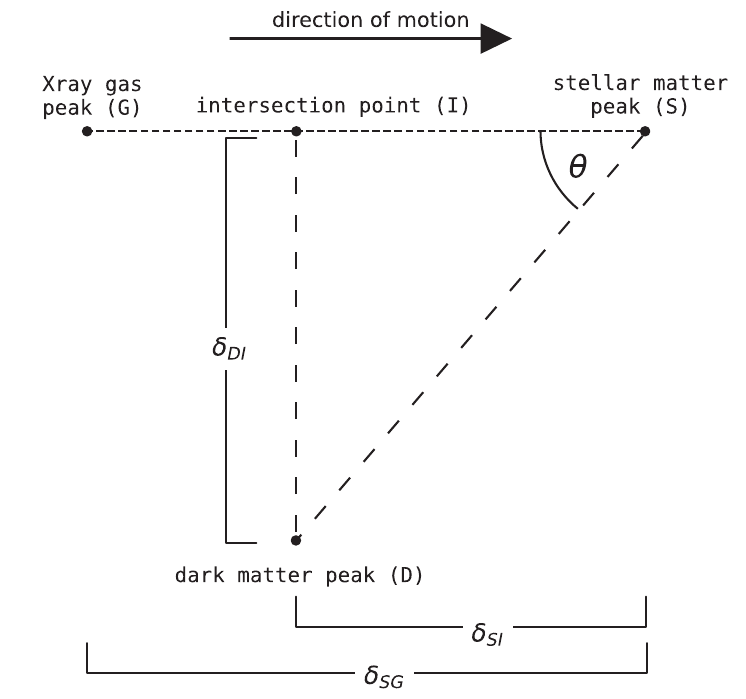}
\caption{During a collision between galaxy clusters, the galaxies (S for `stars'), gas (G), and DM (D) can become separated. Ram pressure on the cluster's gas means that the vector from a cluster's gas to its stars approximately indicates its direction of motion. We define the positive-definite length of this vector $\dSG$. To test whether pressure also acts on DM, we measure the distance from the DM to the stars, in components parallel to the direction of motion, $\dSI$, and perpendicular to it, $\dDI$.}
\label{fig:beta_diagram}
\end{figure}

Consider a triangle (Fig.~\ref{fig:beta_diagram}) with vertices at the centre of mass of stellar matter (S for `stars'), gas (G), and DM (D). The vector from the gas to the stars, $-\rSG$, defines the system's expected direction of motion, and the `base' of the triangle. Whether the locations are defined in 3D or 2D, the spatial offset from the stars to the gas is merely the length of this vector,
\begin{equation}
\dSG = |\rSG|,
\end{equation}
which is positive-definite and has measurement uncertainty 
\begin{equation}
\sSG^2 = \sS^2+\sG^2
\end{equation}
due to uncertainty $\sS$ and $\sG$ in the locations of stellar and gaseous material along a coordinate direction (we assume this to be isotropic).

The location of DM is offset from the location of stars, with components in the direction of motion
\begin{equation}\label{eq:d_si}
\dSI = \frac{\rSG \cdot \rSD}{|\rSG|}~,
\end{equation}
\noindent and perpendicular to it
\begin{equation}\label{eq:d_di}
\dDI = \pm\frac{|\rSG \times \rSD|}{|\rSG|}~,
\end{equation}
where (I) is the point at which a perpendicular from (D) passes through the base of the triangle. Both equation~\eqref{eq:d_si} and \eqref{eq:d_di} can be either positive or negative. In 2D, we take the sign of $\dDI$ to be the sign of the cross-product of $\rSG$ and $\rSD$ (numerator in equation~\eqref{eq:d_di}), resulting in a positive $\dDI$ in Fig.~\ref{fig:beta_diagram}. In 3D, we use the sign of the cross-product of $\rSG$ and $\rS$, dotted with $\rSD$.

Through standard error propagation, these offsets have measurement uncertainty
\begin{equation}
\sSI^2 = \sSD^2 + \frac{2\dSI^2}{\dSG^2}\sSG^2
= \sD^2 + \left(1+\frac{2\dSI^2}{\dSG^2}\right)\sS^2 + \frac{2\dSI^2}{\dSG^2}\sG^2
\end{equation}
and
\begin{equation}
\sDI^2 = \sSD^2 + \frac{2\dDI^2}{\dSG^2}\sSG^2
= \sD^2 + \left(1+\frac{2\dDI^2}{\dSG^2}\right)\sS^2 + \frac{2\dDI^2}{\dSG^2}\sG^2.
\end{equation}

It is informative to calculate the {\it fractional} offset of DM, or `bulleticity'
\begin{equation}\label{eq:beta}
\bpar \equiv \frac{\dSI}{\dSG}.
\end{equation}
This dimensionless ratio has two advantages. First, even though we can observe only the projection of a 3D offset onto the plane of the sky, this quantity is independent of the 3D orientation. Second, an approximate, analytic model of SIDM dynamics suggests that although offsets of each component gradually increase after a merger \citep[see fig.~6 of][]{Robertson_2017a}, the ratio $\bpar$ should be constant for all merger configurations, at all times during the merger. This implies that measurements of $\bpar$ from different systems can be averaged \citep[][we shall discuss this model in more detail in Section~\ref{sec:merging_results}]{Harvey_2015}. 
Measurement noise for individual systems (which may produce $\bpar<0$ or even $\bpar>1$) propagates to uncertainty on $\bpar$ of
\begin{align} \label{eq:sbpar}
\sigma_{\bpar}^{2} &= \frac{1}{\dSG^2}\left[\sSI^2 + \bpar^2\,\sSG^2 \right] \\
&=\frac{1}{\dSG^2}\left[\sD^2 + (1+2\bpar^2)\sS^2 + (2\bpar^2)\sG^2 \right],
\end{align}
where our equation~\eqref{eq:sbpar} recovers equation~(1) of \cite{Wittman_2018}.

\begin{figure*}
\centering
\begin{minipage}{0.48\textwidth}
\includegraphics[width=\linewidth]{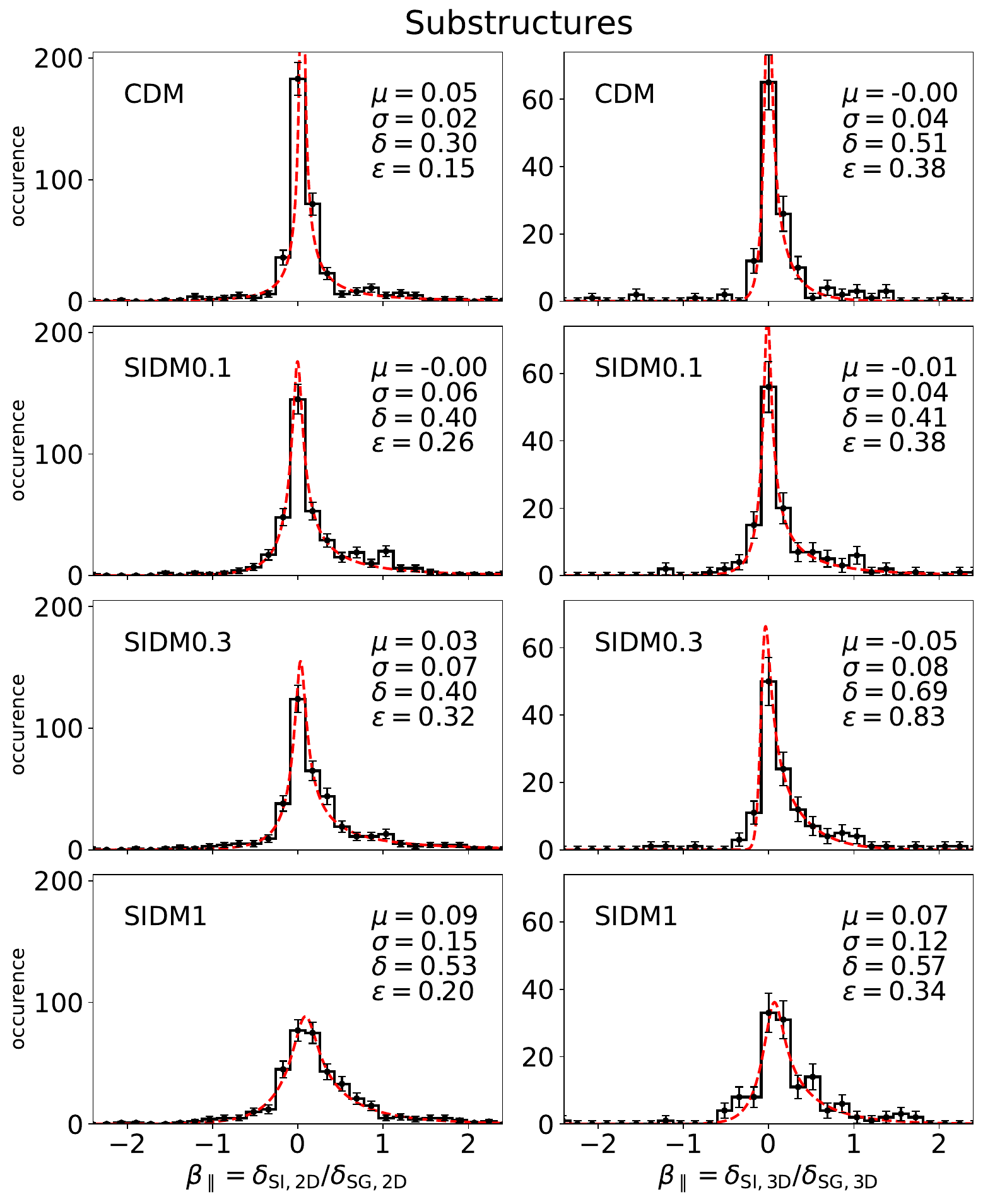}
\end{minipage}\hfill
\begin{minipage}{0.48\textwidth}
\includegraphics[width=\linewidth]{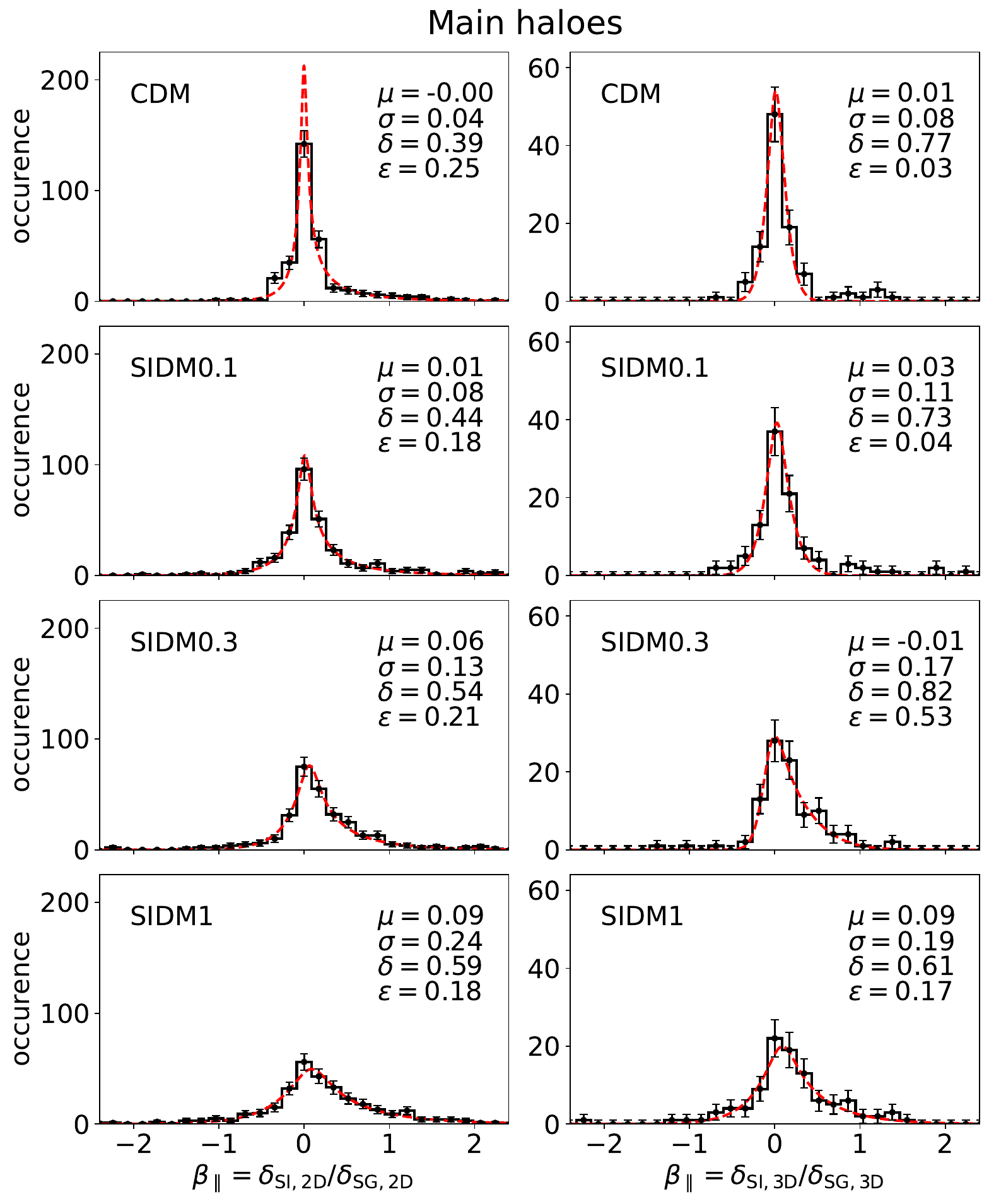}
\end{minipage}
\caption{The fractional offset of DM from galaxies, $\bpar$, in simulations of colliding galaxy clusters. Columns of panels separate the offset in main haloes or subhaloes, and using quantities accessible from 2D projections on the sky or 3D simulations. Rows of panels show results from simulated universes with SIDM cross-section $\sigma/m=[0,\ 0.1,\ 0.3,\ 1]$\,cm$^{2}$\,g$^{-1}$ from top to bottom. Red dashed lines show the best-fitting perturbed Gaussian, see Section~\ref{sec:merging_results}. Fits use the MCMC sampler {\sc emcee}, and assume Poisson noise on each bin.}
\label{fig:distributions_sub}
\end{figure*}

As a control test, we also renormalise the perpendicular offset of DM, 
\begin{equation}\label{eq:betaperp}
\bper \equiv \frac{\dDI}{\dSG},
\end{equation}
which should be consistent with zero on average, if the Universe does not have a handedness (and in the absence of systematics). 
Measurement uncertainty propagates into uncertainty on $\bper$ of
\begin{align} \label{eq:sbper}
\sigma_{\bper}^{2} &= \frac{1}{\dSG^2}\left[\sDI^2 + \bper^2\,\sSG^2 \right] \\
&= \frac{1}{\dSG^2}\left[\sD^2 + (1+2\bper^2)\sS^2 + (2\bper^2)\sG^2 \right].
\end{align}

\subsection{Combining measurements from many collisions}\label{sec:averaging}

If $\bpar$ is universal, it should be possible to measure and interpret the average value $\langle\bpar\rangle$ from a large number of $N_\mathrm{halo}$ merging haloes. Assuming that a given survey will have approximately constant measurement uncertainty $\sD$ and $\sS$, the standard error on the mean of equation~\eqref{eq:beta} is
\begin{equation}\label{eq:snbpar}
\sigma^2_{\langle\bpar\rangle} = \left\langle\frac{1}{\dSG^2}\right\rangle \frac{\sD^2+\sS^2+
2\bpar^2(\sS^2+\sG^2)
}{N_\mathrm{halo}}
.
\end{equation}

\begin{figure}
\centering
\includegraphics[width=\linewidth]{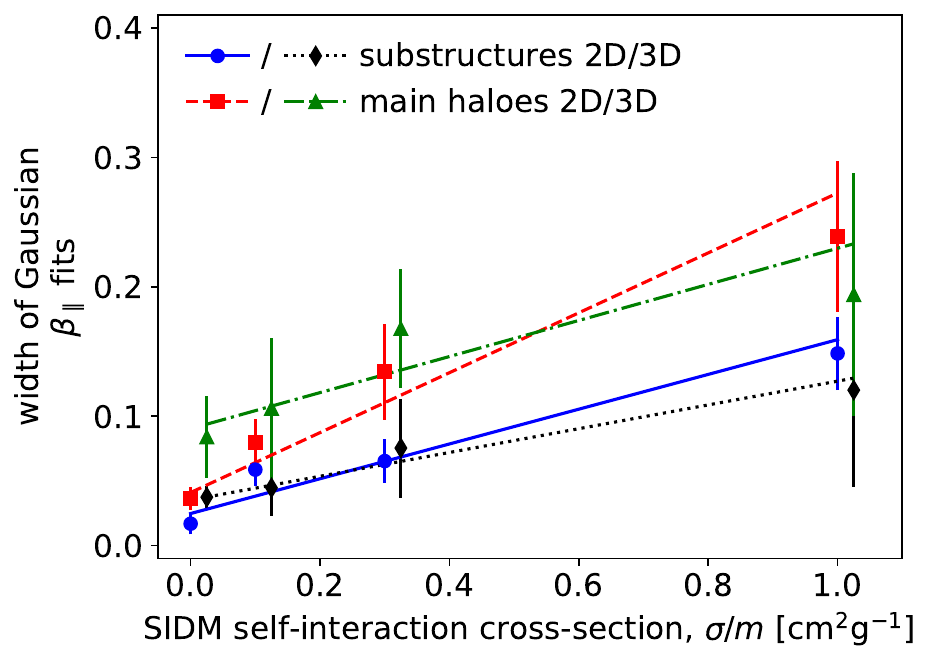}
\caption{The width of the best-fitting perturbed Gaussian to distributions of $\bpar$ in Fig.~\ref{fig:distributions_sub}.
Variation between clusters grows with SIDM cross-section for main haloes (red squares and dashed line for 2D, green triangles and dash-dotted lined for 3D) and subhaloes (blue circles and solid line for 2D, black diamonds and dotted line for 3D). The errors are the 1$\sigma$ uncertainties returned by the fit to the distributions in Fig.~\ref{fig:distributions_sub}. The straight lines plotted here were fitted to the data points using the Python function \textsc{scipy.optimize.curve\_fit}. The 3D data points have been slightly offset for clarity.}
\label{fig:width_vs_sigma}
\end{figure}

Some merging systems have more discriminating power than others \citep{Wittman_2018}. A measurement of $\bpar$ is a calibration of the location of DM, some distance $\dSI$ along a ruler of length $\dSG$. If measurement precision is constant, systems with a long ruler offer high dynamic range and high signal-to-noise. Measurement precision is roughly constant in terms of angle on the sky, so systems with the longest rulers are those near the viewer, those with a collision aligned in the plane of the sky, and those with timing such that the separation is maximised. When analysing observations, \cite{Wittman_2018} found it helpful to average systems using an inverse-variance weight
\begin{equation}\label{eq:weightapprox}
w_{i} \propto \delta^2_{\mathrm{SG},i}~, 
\end{equation}
which is obtained from equation~\eqref{eq:snbpar} while ignoring terms $\mathcal{O}(\bpar^2)$ both because they are small, and to avoid biasing the measurement of $\bpar$ itself. The error on the weighted mean then becomes
\begin{equation}\label{eq:snwbpar}
\sigma^2_{\langle\bpar\rangle_w} = \frac{1}{\left\langle\dSG^2\right\rangle} \frac{\sD^2+\sS^2+\mathcal{O}(\bpar^2)}{N_\mathrm{halo}}~.
\end{equation}

In this paper, we shall use neither means nor weighted means. 
We shall instead quote measurements of $\bpar$ using a median, and `$1\sigma$-like' uncertainties (separation between the $16^\mathrm{th}$ and $84^\mathrm{th}$ percentiles of the distribution). The median is similar to the mean within scatter, but in simulations rather than observations, our automated methods produce unstable scatter in the mean -- and even more scatter in the weighted mean.
A small fraction of the time this is because the shrinking spheres method failed catastrophically (see Section~\ref{sec:shrinking_spheres}); 
it is difficult to exclude failed measurements via cuts on $\dSG$, because those cuts would exclude haloes with the highest signal to noise -- and the weight becomes strongly dominated by the system with the largest $\dSG$ that survives the cut. 
For example, a single subhalo with large $\dSG$ in our sample increases the weighted mean by a factor of nearly 10 compared to the value if it is excluded. A weighted mean might be more suitable when clean measurements are available, i.e.\ when random statistical errors dominate over systematic errors. 
Curiously, we find that our results are stable when using weight $w_{i} \propto \delta_\mathrm{SG,i}$ (see Appendix~\ref{sec:w_median}). This is not motivated mathematically, but empirically we find that it would be worth investigating in the future.

\section{Results}\label{sec:merging_results}

Measurements of $\bpar$ from simulated merging clusters follow an approximately Gaussian distribution with an asymmetric tail to positive values (see Fig.~\ref{fig:distributions_sub}, and Appendix~\ref{sec:indiv_components} for measurements of individual components), aside from the outliers discussed above. Measurements of $\bper$ are similar but without the tail. Distributions of $\bpar$ are remarkably consistent between subhaloes (leftmost two columns in Fig.~\ref{fig:distributions_sub}) and main haloes (rightmost two columns), and almost indistinguishable whether measured in 2D (first and third columns) or 3D (second and fourth columns). We fit a Gaussian perturbed with skewness $\epsilon$ and kurtosis $\delta$ \citep[the sinh-arcsinh normal distribution,][]{Jones_2009}
\begin{equation}\label{eq:skew_Gaussian}
f(x;\mu,\sigma,\delta,\epsilon) = \frac{\delta}{\sigma} \sqrt{\frac{1+S^{2}(y;\delta,\epsilon)}{2\pi(1+y^{2})}}\exp\left(-\frac{1}{2}S^{2}(y;\delta,\epsilon)\right)
\end{equation}
to all of these profiles, where $S(y;\delta,\epsilon) = \sinh(\delta \sinh^{-1}(y) - \epsilon)$ and $y=(x-\mu)/\sigma$. When $\epsilon=0$ and $\delta=1$, this reduces to the normal distribution with mean $\mu$ and standard deviation $\sigma$. We average this analytic function in the same bins as the data (30 bins of equal width between -2.5 and 2.5), then use the Markov Chain Monte Carlo (MCMC) sampler {\sc emcee} \citep{Emcee_2013} to obtain the maximum-likelihood values and posterior PDFs of the free parameters.

\subsection{Scatter of DM offsets}

We first recover the result noticed by \cite{Kim_2017} and \cite{Harvey_2017, harvey19} that scatter in measurements of $\bpar$ increases with SIDM cross-section $\sigma/m$ (see Fig.~\ref{fig:width_vs_sigma}; the same is true for $\bper$). This is likely due to the collision giving an impulse to the BCG that sets it oscillating. Since it oscillates within a gravitational potential that is dominated by DM, and SIDM clusters have a `core' of constant density, the BCG is less tightly bound and its oscillations have a larger amplitude. The observed scatter in offsets samples random phases of oscillation at the moment when it is measured.

We discover that the scatter of $\bpar$ and $\bper$ in subhaloes also increases with SIDM cross-section $\sigma/m$, although less than that in main haloes. 
This may be simply because subhaloes are tidally disrupted before their offsets increase to the same extent as main haloes.
Scatter in offsets could be used to measure the SIDM cross-section in the real Universe \citep{harvey19}. However, there is no null test for CDM, and no control test for systematics 
-- so its interpretation will rely entirely on calibration via simulations. We shall not consider it further in this paper.

\begin{figure}
\centering
\includegraphics[width=\linewidth,trim={0mm 3mm 0mm 0mm,clip}]{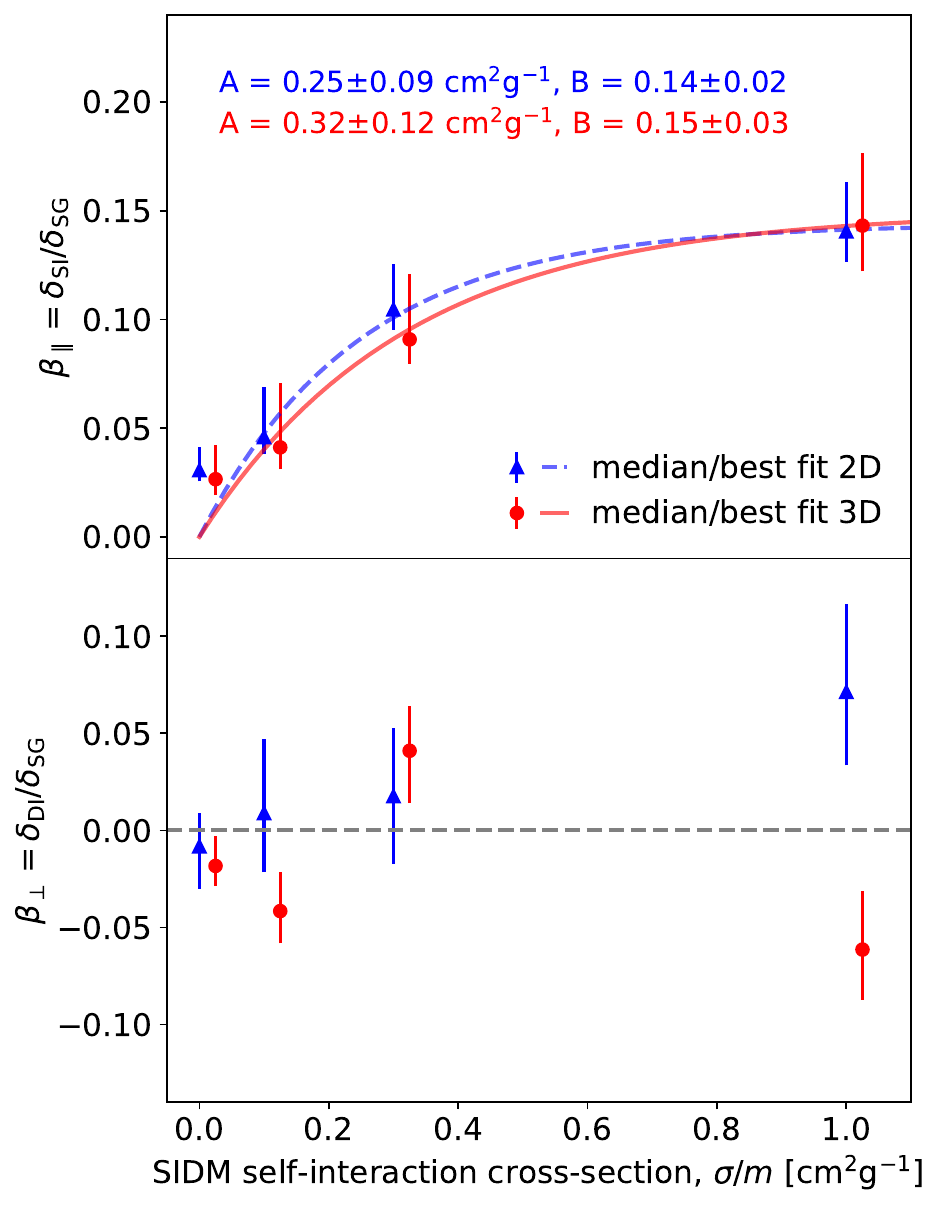}
\caption{Median fractional offset between galaxies and DM in simulated colliding clusters, as a function of the SIDM cross-section $\sigma/m=[0,\ 0.1,\ 0.3,\ 1]$\,cm$^{2}$\,g$^{-1}$
(top panel), and the perpendicular control test that should be consistent with zero in the absence of systematics (bottom panel). Red circles and blue triangles show similar calculations using information available in 3D or that projected onto a 2D sky; for clarity, 3D data points are horizontally offset by a small amount.
Error bars show `$1\sigma$-like' uncertainty, between the $16^\mathrm{th}$ and $84^\mathrm{th}$ percentiles.
Curves show the best fits of model~\eqref{eq:beta_fit_func}, with parameters tabulated in the bottom row of Table~\ref{tab:fit_params}.
}
\label{fig:beta_sim_all}
\end{figure}

\subsection{Typical DM offset}
\label{sec:typical_offset}

The median value of $\bpar$ increases with SIDM cross-section $\sigma/m$. Measurements are generally consistent for both main haloes and subhaloes, so we show the median of the combined sample (Fig.~\ref{fig:beta_sim_all}).
Measurements of the perpendicular control test, $\bper$, are consistent with zero as expected.
Importantly, results are virtually indistinguishable whether quantities are calculated in 3D after projection into 2D. In this sense, the measurement should therefore be as accessible to observations as it is to simulations.

Our measurements are remarkably well fit by the model 
\begin{equation}\label{eq:beta_fit_func}
\bpar(\sigma/m) = B\left(1-e^{-(\sigma/m)/A}\right)
\end{equation}
predicted by \citet[][see their equation~33 and fig.~2]{Harvey_2014}. This model interpolates between two well-understood extremes. At low $\sigma$, the halo is optically thin, and the effective drag force grows linearly with the interaction cross-section. The constant of proportionality $A$ reflects the relative interaction strengths of DM and gas (or the characteristic cross-section at which a halo becomes optically thick). At high $\sigma$, DM particles at the front of the halo always scatter incoming DM and shield particles behind, so the halo becomes optically thick. The drag then becomes a constant, with parameter $B$ that depends on the geometry of the halo. The model predicts that $\bpar$ is notably independent of infall velocity, impact parameter, and time. We fit free parameters $A$ and $B$ to our data using the Python function \textsc{scipy.optimize.minimize} and an asymmetric loss function to account for asymmetric errors bars (Table~\ref{tab:fit_params}, with best-fitting models overlaid in Fig.~\ref{fig:beta_sim_all}).

Our only measurement not well fit by model~\eqref{eq:beta_fit_func} is that, for subhaloes in CDM ($\sigma/m=0$) simulations, we measure median $\bpar>0$ at $\sim$95~per~cent confidence level. That we would not expect CDM to be offset from galaxies is particularly important as a null test. With our present sample size, we tentatively ascribe this as a statistical fluke or a limitation of simulation resolution (for $\dSG\approx 23$~kpc, a value $\bpar=0.03$ represents a measurement of $\dSI$ to ${\sim}$12~per~cent of the BAHAMAS softening length). We believe this because our measurements of \textit{higher mass} CDM main haloes (not shown by themselves) are consistent with zero offset, and \cite{schaller15}'s measurements of \textit{lower mass} individual galaxies also show zero offset in CDM simulations with better resolution. Furthermore, the mean (rather than median) value of $\bpar$ happens to be slightly positive for main haloes and slightly negative for subhaloes. Nonetheless, this should be re-measured in future work. \cite{Robertson_2017a} measured no offset in idealised simulations, so it is feasible that this is a new effect particular to fully cosmological simulations, caused by the varied impact parameters, ongoing star formation, or chance projection of substructures (which dominate measurement noise; see Section~\ref{sec:shrinking_spheres}).

Finally, it is interesting to note that \cite{Robertson_2017a} report a small bias when measuring the position of one halo in the outskirts of another, because the second halo contributes a gradient of particles across the shrinking circle. We confirm this, by comparing positions measured (by default) using all particles, with positions measured using only bound particles. We find that positions move by a comparable amount in the same direction, such that the effect on $\bpar$ is negligible. Using angle brackets to denote medians, we measure in CDM simulations a decrease from $\langle\bpar\rangle^\mathrm{all}_\mathrm{2D}=0.0307^{+0.011}_{-0.005}$ to $\langle\bpar\rangle^\mathrm{bound}_\mathrm{2D}=0.0302^{+0.011}_{-0.006}$, and an increase from $\langle\bpar\rangle^\mathrm{all}_\mathrm{3D}=0.0265^{+0.016}_{-0.007}$ to $\langle\bpar\rangle^\mathrm{bound}_\mathrm{3D}=0.0273^{+0.017}_{-0.008}$.

\begin{table}
\centering
\caption{Best-fitting parameters of \protect\cite{Harvey_2014}'s analytic model $\bpar(\sigma/m)$ (equation~\ref{eq:beta_fit_func}) to our measurements of median $\bpar$, from quantities accessible in either 2D or 3D. Rows show results from the entire sample, just the main haloes, just the substructures, and haloes in 2-body systems, split by the sign of their relative velocity (i.e.\ whether they are approaching pericentre or receding after it).}
\label{tab:fit_params}
\begin{tabular*}{0.95\columnwidth}{lcccc}
\hline
& \multicolumn{2}{c}{2D} & \multicolumn{2}{c}{3D} \\\hline
& $A$ & $B$ & $A$ & $B$ \\
& [$\mathrm{cm^{2}\,g^{-1}}$] & & [$\mathrm{cm^{2}\,g^{-1}}$] & \\\hline
All haloes & 0.25$\pm$0.09 & 0.14$\pm$0.02 & 0.32$\pm$0.12 & 0.15$\pm$0.03\\
Main haloes & 0.28$\pm$0.13 & 0.14$\pm$0.03 & 0.27$\pm$0.09 & 0.14$\pm$0.02\\
Substructures & 0.36$\pm$0.16 & 0.15$\pm$0.03 & 0.42$\pm$0.19 & 0.16$\pm$0.04\\
Approaching & 0.11$\pm$0.14 & 0.12$\pm$0.13 & 0.13$\pm$0.04 & 0.24$\pm$0.13\\
Receding & 0.19$\pm$0.11 & 0.36$\pm$0.38 & 0.14$\pm$0.05 & 0.10$\pm$0.09\\
\hline
\end{tabular*}
\end{table}

\subsection{Selection effects}

Our interpretation of the average measurement from many colliding systems presupposes that $\langle\bpar\rangle$ is universal (see the start of Section~\ref{sec:averaging}). The components of our simulated mergers are typically closer than those in observations (see Section~\ref{sec:merging_sample}). Furthermore, the offsets of individual components of matter vary with time since collision \citep{Robertson_2017a}. If $\bpar$ also varies with time, it matters whether clusters are observed before or after pericentric passage.

Unfortunately, the BAHAMAS simulations do not include enough cluster mergers to reach discriminating statistics if we split our sample, nor enough snapshots to measure them at different times since collision. Indeed, it is not always clear from the single snapshot available whether clusters have already reached pericentre. For systems with only two haloes, splitting by whether the net velocity of their particles is approaching or receding yields unstable measurements of parameter $A$ and much larger uncertainties in parameter $B$ (Table~\ref{tab:fit_params}). For two-halo systems in SIDM1 simulations analysed in 2D (using angle brackets to denote medians, and propagating 16/84 percentile uncertainties like standard deviations), we find $\langle\bpar\rangle^\mathrm{recede}/\langle\bpar\rangle^\mathrm{approach}=1.39^{+0.62}_{-0.46}$ (or $0.39^{+0.48}_{-0.30}$ split further into main haloes, and $2.84^{+3.17}_{-0.87}$ for subhaloes). 
Future investigation of this would be interesting with larger simulations. If it makes a significant difference to predictions of $\langle\bpar\rangle$, the selection of simulated clusters should be matched to selection biases in observational samples. Alternatively, observational samples could be selected carefully, inferring the direction of motion using a combination of optical and X-ray data, or shock fronts in e.g.\ radio emission.

\subsection{Future prospects}

Measurements of $\bpar$ are a promising way to test the interaction cross-section of DM. Symmetries are expected to provide a null result $\bpar=0$ in the case of CDM ($\soverm=0$) and a perpendicular test for systematics, $\bper$. Future surveys may be able to combine measurements of $\bpar$ and $\bper$ from a large number of observed merging galaxy clusters. 

\begin{figure}
\centering
\includegraphics[width=\linewidth,trim={0mm 0mm 0mm 0mm,clip}]{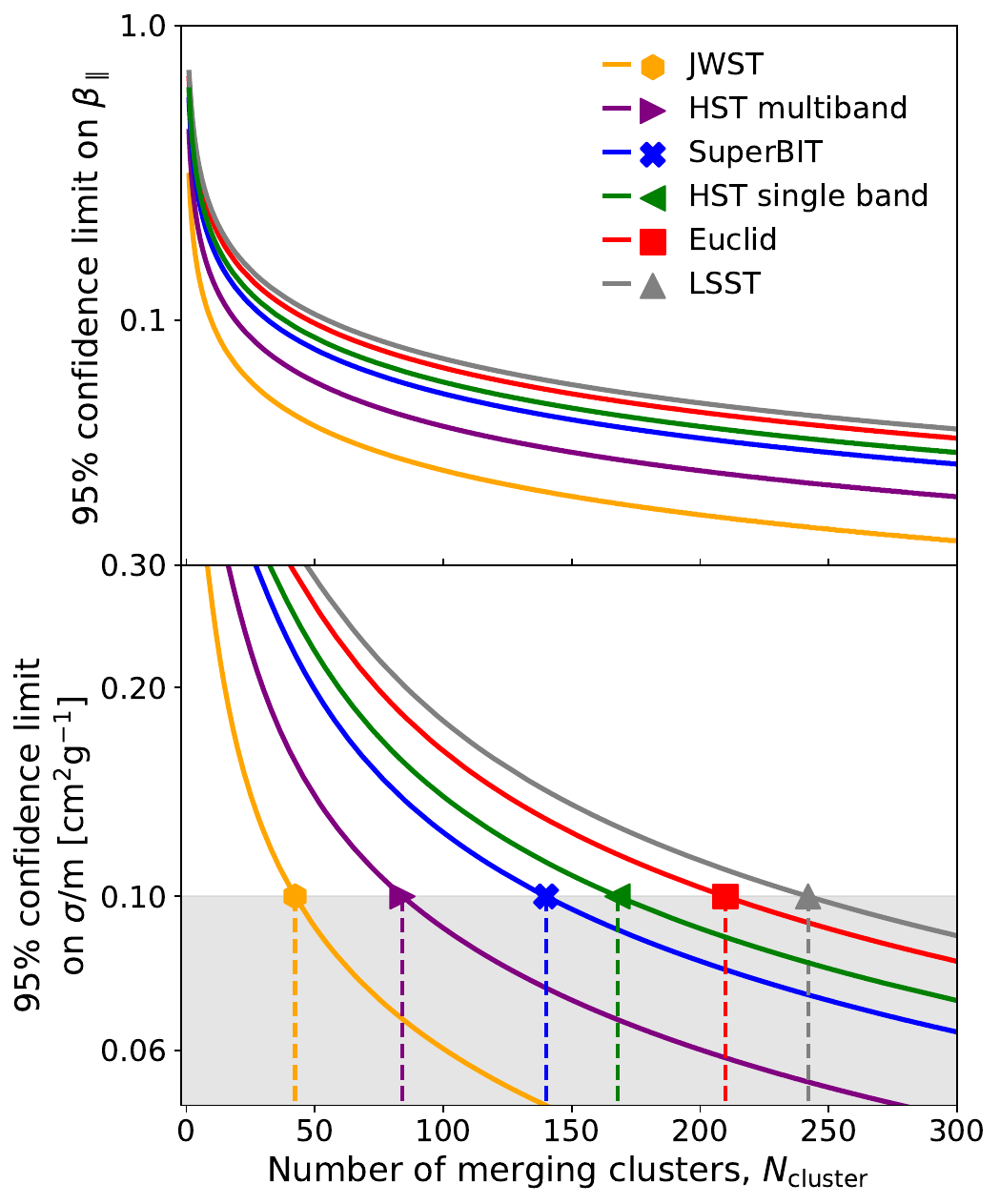}
\caption{Forecast upper 95~per~cent confidence limits on $\bpar$ (top) and $\soverm$ (bottom) for future observations with JWST (gold hexagon), HST single- and multi-band imaging (green triangle pointing left and purple triangle pointing right), SuperBIT (blue cross), Euclid (red square) and LSST (grey triangle pointing up), as a function of the number of merging clusters observed.
Most predictions assume multicolour imaging; this experiment has only ever been performed in practice with single-band HST imaging.}
\label{fig:superbit_pred}
\end{figure}

To forecast a future survey's ability to rule out the null hypothesis $\soverm=0$, we first consider the expected uncertainty on $\langle\bpar\rangle_{w}$ (equation~\ref{eq:snwbpar}), setting $\sigma/m=0$, and bootstrapping the other parameters from observational experience with single-band HST data \citep{Harvey_2015}.
A large sample of merging clusters should at least maintain \citet{Harvey_2015}'s sample mean\footnote{The relevant quantity for unweighted averages (equation~\ref{eq:snbpar}) is $\langle 1/\dSG^{2}\rangle=85.2\times10^{-3}$\,arcsec$^{-2}$, although we do not use it here.} of $1/\langle\dSG^{2}\rangle=1.6\times10^{-3}$\,arcsec$^{-2}$ (as discussed in Section~\ref{sec:merging_sample}, but converting the number into angular units), with $N_\mathrm{halo}=2.3N_\mathrm{cluster}$. 
The observationally achieved uncertainty on the location of stellar material, $\sS=0.6$~arcsec, is both subdominant and unlikely to change significantly because it depends on astrophysical effects such as confusion between multiple BCGs \citep{George_2012}. Analysis of mock images suggests the achieved uncertainty on the location of DM, $\sD=11.4$~arcsec, is reduced by approximately 30~per~cent by perfect separation between galaxies in front of or behind the cluster using multicolour photometry, and falls proportionally to $1/\sqrt{\ngal}$, the density of resolved background galaxies \citep{Harvey_2013}.
Table~\ref{tab:future_surveys} collates expectations of $\ngal$ for the James Webb Space Telescope \citep[JWST,][]{Gardner_2006,cosmos_webb}, Super-Pressure Balloon-borne Imaging Telescope \citep[SuperBIT,][]{Romualdez_2016,Shaaban_22}, Euclid \citep{Euclid_2011,Euclid_wide} and the Vera C.\ Rubin Observatory \citep[LSST,][]{Chang_2013,LSST_2019}.

For each future survey we calculate the single-tailed 95~per~cent confidence limit on
\begin{equation}\label{eq:snwbparv}
\sigma^2_{\langle\bpar\rangle_w} = \frac{\pi}{2}\times
\frac{1.6\times10^{-3}}{2.3\,N_\mathrm{cluster}}\times\left(\sS^{2}+\sD^{2}\right)~,
\end{equation}
using the values in Table~\ref{tab:future_surveys} and assuming a Gaussian error distribution \citep[which Fig.~3B of][suggests to be reasonable]{Harvey_2015}.We estimate the standard error on the median by multiplying the standard error of the mean by a factor of $\sqrt{{\pi}/{2}}$ \citep[for details see chapter~4 of][]{Maindonald_Braun_2010}. We finally convert this to a single-tailed 95~per~cent confidence limit on $\sigma/m$ using equation~\eqref{eq:beta_fit_func} with parameter values from Table~\ref{tab:fit_params} (all haloes, 2D). These two confidence limits are shown, as a function of the number of observed systems, in Fig.~\ref{fig:superbit_pred}. The number of typical merging clusters that must be observed by any telescope to reach the target particle physics sensitivity of 0.1\,$\mathrm{cm^{2}\,g^{-1}}$ can be read from the bottom panel of Fig.~\ref{fig:superbit_pred}, and is listed in Table~\ref{tab:future_surveys}.
Indeed, SuperBIT's design characteristics were optimised to meet this goal as its primary science driver \citep{McCleary_2023}.

\begin{table}
\centering
\caption{Assumed characteristics of hypothetical astronomical observations that could be used to measure the offset of DM from ordinary matter, as proposed in this paper. For various telescopes, columns indicate: the density of resolved galaxies behind merging clusters, $\ngal$, the precision with which it is possible to measure the location of stellar material, $\sS$, and DM, $\sD$. The final column indicates the number of clusters, $N_\mathrm{cluster}$, that must be observed to potentially rule out the hypothesis $\soverm=0$ with 95~per~cent confidence.\!
\label{tab:future_surveys}
}
\begin{tabular*}{0.95\columnwidth}{lccr@{$/\sqrt{\ngal}=$}r@.lr}
\hline
\multirow{2}{*}{Telescope} & $\ngal$ & $\sS$ & \multicolumn{3}{c}{$\sD$} & \!\!\!\!\!\!Required \\
 & [arcmin$^{-2}$] & \!\!\![arcsec]\!\!\! & \multicolumn{3}{c}{[arcsec]} & \!\!\!$N_\mathrm{cluster}$~\\
 \hline
JWST & $150$ & $0.6$ & $69.8$ & 5&7 & 42~~~\\
HST (multi-band)\hspace{-5mm} & $75$ & $0.6$ & $69.8$ & 8&1 & 84~~~\\
~~~~~~~~\,(single band)\hspace{-5mm} & $75$ & $0.6$ & $98.7$ & 11&4 & 168~~ \\
SuperBIT & $40$ & $0.6$ & $69.8$ & 10&4 & 140~~~\\
Euclid & $30$ & $0.6$ & $69.8$ & 12&7 & 210~~~\\
LSST & $26$ & $0.6$ & $69.8$ & 13&7 & 242~~~\\
\hline
\end{tabular*}
\end{table}

This forecast is valid for surveys measuring (or placing upper limits on) SIDM cross-section $0.05\lesssim\sigma/m\lesssim 0.5\,\mathrm{cm^{2}\,g^{-1}}$. At lower cross-sections, either simulation resolution, noise, or cosmological effects (see Section~\ref{sec:typical_offset}) inhibit simulated values of $\bpar$ reaching zero. Such tight constraints would require observations of approximately three times more clusters than listed in Table~\ref{tab:future_surveys}. At higher cross-sections, the limited number of simulated clusters leads to measurement uncertainty such that sometimes $\langle\bpar\rangle>B$, which is incompatible with equation~\eqref{eq:beta_fit_func}. 
That model cannot be used to map $\langle\bpar\rangle$ onto $\soverm$, 
until sufficient clusters have been observed such that (in this case) 95~per~cent of the posterior probability is at values of $\langle\bpar\rangle<B$.

\section{Conclusions}\label{sec:conclusions}

Terrestrial particle physics experiments have established that DM interacts very weakly, if at all, with Standard Model particles. 
However, terrestrial experiments are unable to test whether DM particles interact with {\it each other} -- as predicted for many proposed models of self-interacting DM \citep[SIDM, e.g.][]{sidm_model_qball,sidm_model_axion,sidm_model_yukawa,maxSIDM}, including a large class containing a dark-sector analogue of the strong force \citep{sidm_model_mirror,sidm_model_mirror2,sidm_model_hiddensector}. 
In the latter models, the natural scale of the interaction cross-section is the same order as for nuclear interactions, $\sigma/m\sim 0.6$\,cm$^{2}$\,g$^{-1}$. 

Using cosmological simulations, we have measured the effect of DM self-interactions on the major mergers of galaxy clusters. We find that the offset between DM and galaxies (as a fraction of that between gas and galaxies) is a promising test of SIDM. This `bullet\-icity', $\bpar$, increases with cross-section in a way that matches the predictions from an analytic model (equation~\ref{eq:beta_fit_func}) originally proposed by \citet{Harvey_2014}. Because it is a fractional offset, the same measurements can be accessed using either 3D or 2D projected data. Symmetries provide a null test $\langle\bpar\rangle=0$ for non-interacting DM, and an orthogonal test $\langle\bper\rangle=0$ for systematics or to measure scatter. 

Three challenges remain with theoretical predictions. 
First, because of instabilities in our identification of peak positions, we find the median offset of DM haloes to be more robust than the mean. It would be interesting to repeat this analysis with methods for peak finding that are closer to those used with observational data. These may differently weight the distribution of matter close to a peak or at distance from it, which is important if the distribution is skewed. Second, we find that the median (and mean) $\bpar$ is slightly positive for all values of interaction cross-section, even $\sigma/m=0$ (Fig.~\ref{fig:beta_sim_all}). This implies that it is more likely for DM to be found between the stars and gas, rather than leading the stars. Other CDM simulations \citep{schaller15, Robertson_2017a}, and an analytic model (equation~\ref{eq:beta_fit_func}) predicted $\langle\bpar\rangle$ to be consistent with zero when $\sigma/m=0$. 
Our measurement of an offset in the non-interacting CDM case might be a statistical anomaly, might be caused by the matching of DM to galaxy and gas peaks when starting shrinking spheres from {\sc subfind} positions, or it might be a symptom of more complex dynamical processes. Third, our simulated survey volume is too small to contain sufficient systems for the sample to be usefully split by e.g.\ mass ratio, impact speed, or time before/after first pericentric passage. It will be important to test whether $\bpar$ is a universal function of $\sigma/m$ for a range of these parameters, as predicted. We have no evidence from BAHAMAS to indicate that it is not, but are running larger simulation volumes precisely to test this. In future simulations, it will also be interesting to measure the $\bpar$ produced by SIDM interactions mediated by low-mass particles that produce more frequent but smaller momentum exchange scattering events, which manifests on macroscopic scales as something closer to a drag force \citep{Fischer_2022}.

Finally, we made predictions for limits on the DM self-interaction cross-section that could be ascertained by future telescopes. The test is promising, with astronomical observations of $\sim$$100$ merging clusters yielding constraints relevant to particle physics. If $\bpar$ is not perfectly universal, larger simulations will also be important to interpret observations, by reproducing sample selection effects -- e.g.\ with larger values of $\langle\dSG\rangle$ and more systems just after first pericentric passage. Reproducing selection effects would be even more important if averaging samples with a mean or weighted mean rather than a median, because these are so strongly dominated by a small number of systems.

\section*{Acknowledgements}
We thank Adam Amara and Mathilde Jauzac for encouraging feedback on an early draft of this paper, Ian McCarthy for sharing his CDM simulation runs, and the anonymous referee for helpful suggestions that improved our analysis. ELS was supported by the Australian Government through the Australian Research Council Centre of Excellence for Dark Matter Particle Physics (CDM, CE200100008) and by the Royal Society. RJM was supported by the Royal Society. KAO acknowledges support support by STFC through grant ST/T000244/1. KAO and CSF were supported by the European Research Council (ERC) through Advanced Investigator grant DMIDAS (GA 786910).
This work used the DiRAC@Durham facility managed by the Institute for Computational Cosmology on behalf of the STFC DiRAC HPC Facility (www.dirac.ac.uk). The equipment was funded by BEIS via STFC capital grants ST/K00042X/1, ST/P002293/1, ST/R002371/1 and ST/S002502/1, Durham University and STFC operations grant ST/R000832/1. DiRAC is part of the UK National e-Infrastructure. 

\section*{Data Availability}
The CDM simulations were introduced by \cite{McCarthy_2017}, and the SIDM simulations by \cite{Robertson_2017b}. All simulation data used in this paper are available via those original papers.

\bibliographystyle{mnras}
\bibliography{references}

\appendix

\section{Weighted median of ensemble}\label{sec:w_median}

We here investigate the weighted median of $\bpar$. The weighted median is equal to the weighted 50$^\mathrm{th}$ percentile, where the weighted 100$p^\mathrm{th}$ percentile ($0 < p < 1$) is calculated by sorting the data and finding the smallest set of data for which the weights sum to a fraction $p$ of the total weight.

\begin{figure}
\centering
\includegraphics[width=0.991\linewidth,trim={0mm 0mm 0mm 0mm,clip}]{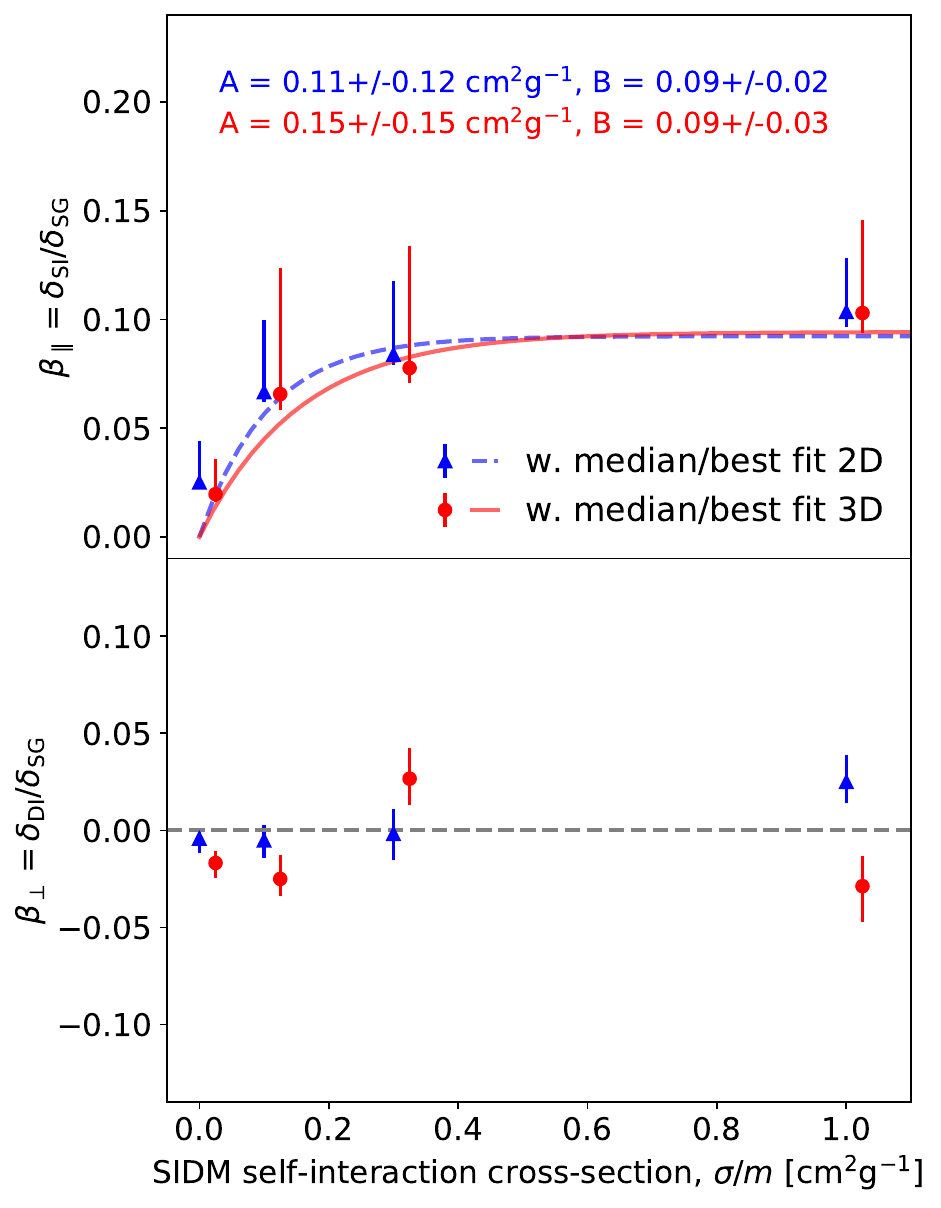}
\caption{Weighted median fractional offset between galaxies and DM in simulated colliding clusters, as a function of the SIDM cross-section $\sigma/m=[0,\ 0.1,\ 0.3,\ 1]$\,cm$^2$g$^{-1}$
(top panel), and the perpendicular control test that should be consistent with zero in the absence of systematics (bottom panel). Red circles and blue triangles show similar calculations using information available in 3D or that projected onto a 2D sky; for clarity, 3D data points are horizontally offset by a small amount.
Error bars show `$1\sigma$-like' uncertainty, i.e.\ the separation between the 16$^\mathrm{th}$ and 84$^\mathrm{th}$ percentiles of the distributions.
Curves show the best fits of model~\eqref{eq:beta_fit_func}, with parameters tabulated in the bottom row of Table~\ref{tab:fit_params_w_median}.
}
\label{fig:beta_sim_all_w}
\end{figure}

Using inverse variance weights $w_{i}\propto\delta_\mathrm{SG, i}^{2}$ (Equation~\ref{eq:weightapprox}), we found that the median $\bpar$ becomes highly unstable. Curiously, it {\it is} stable with weight $w_{i}\propto\delta_\mathrm{SG, i}$ (Fig.~\ref{fig:beta_sim_all_w} and Table~\ref{tab:fit_params_w_median}). This is because it calculates the median of $\langle\dSI\rangle/\langle\dSG\rangle$. This scheme is unjustified mathematically, but would be worth considering in future analyses.

\begin{table}~~~
\centering
\caption{Best-fitting parameters of \protect\cite{Harvey_2014}'s analytic model $\bpar(\sigma/m)$ (equation~\ref{eq:beta_fit_func}) to our measurements of ($\dSG$-)weighted median $\bpar$, from quantities accessible in either 2D or 3D. Rows show results from just the substructures, just the main haloes, and everything combined.}
\label{tab:fit_params_w_median}
\begin{tabular*}{0.95\columnwidth}{lcccc}
\hline
& \multicolumn{2}{c}{2D} & \multicolumn{2}{c}{3D} \\\hline
& $A$ & $B$ & $A$ & $B$ \\
& [$\mathrm{cm^{2}\,g^{-1}}$] & & [$\mathrm{cm^{2}\,g^{-1}}$] & \\\hline
All haloes & 0.11$\pm$0.12 & 0.09$\pm$0.02 & 0.15$\pm$0.15 & 0.09$\pm$0.03\\
Main haloes & 0.02$\pm$0.05 & 0.09$\pm$0.01 & 0.05$\pm$0.02 & 0.09$\pm$0.01\\
Substructures & 0.48$\pm$0.19 & 0.16$\pm$0.04 & 0.63$\pm$0.38 & 0.16$\pm$0.06\\
\hline
\end{tabular*}
\end{table}

\section{Offsets of individual components}\label{sec:indiv_components}

\begin{figure*}
\centering
\begin{minipage}{0.48\textwidth}
\includegraphics[width=\linewidth]{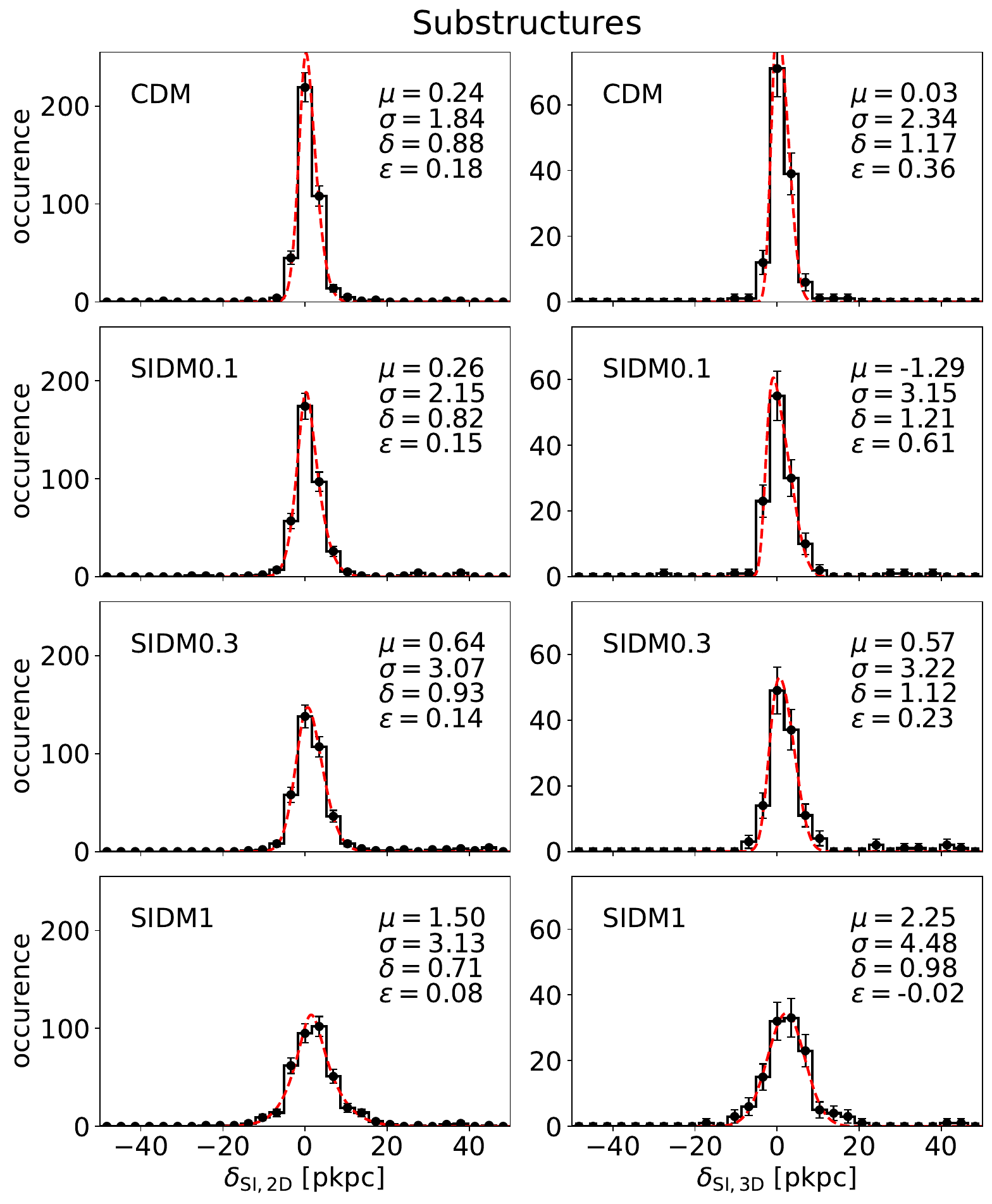}
\end{minipage}\hfill
\begin{minipage}{0.48\textwidth}
\includegraphics[width=\linewidth]{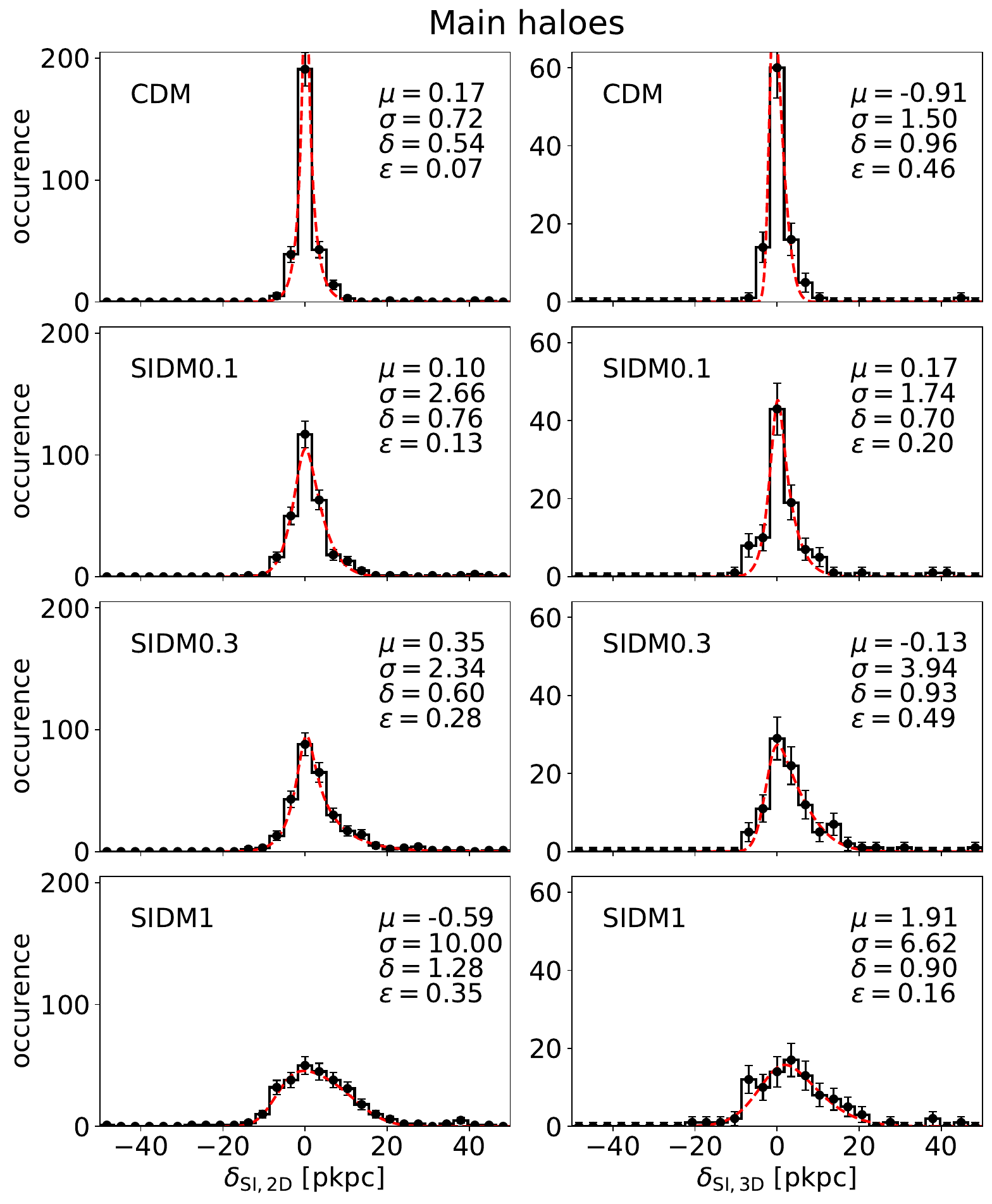}
\end{minipage}
\caption{The distance between stellar matter and DM in simulations of colliding galaxy clusters, {\it parallel} to the direction of motion (see Fig.~\ref{fig:beta_diagram}). Columns of panels separate the distance in main haloes or subhaloes, and using quantities accessible from 2D projections on the sky or 3D simulations. Rows of panels show results from simulated universes with SIDM cross-section $\sigma/m=[0,\ 0.1,\ 0.3,\ 1]$\,cm$^2$g$^{-1}$ from top to bottom. Note that the $y$-axis range is different in each column.}
\label{fig:si_distributions}
\end{figure*}

\begin{figure*}
\centering
\begin{minipage}{0.48\textwidth}
\includegraphics[width=\linewidth]{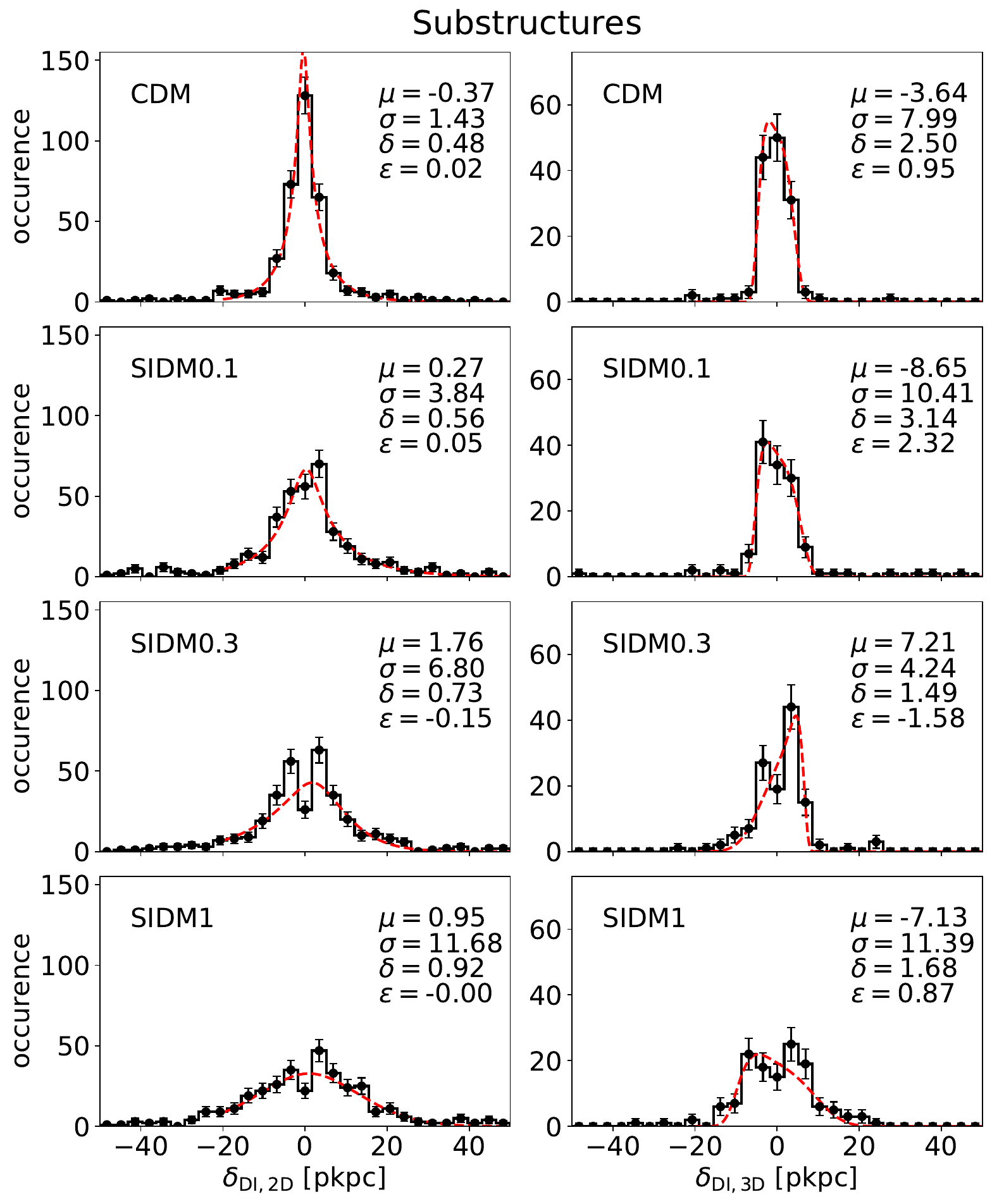}
\end{minipage}\hfill
\begin{minipage}{0.48\textwidth}
\includegraphics[width=\linewidth]{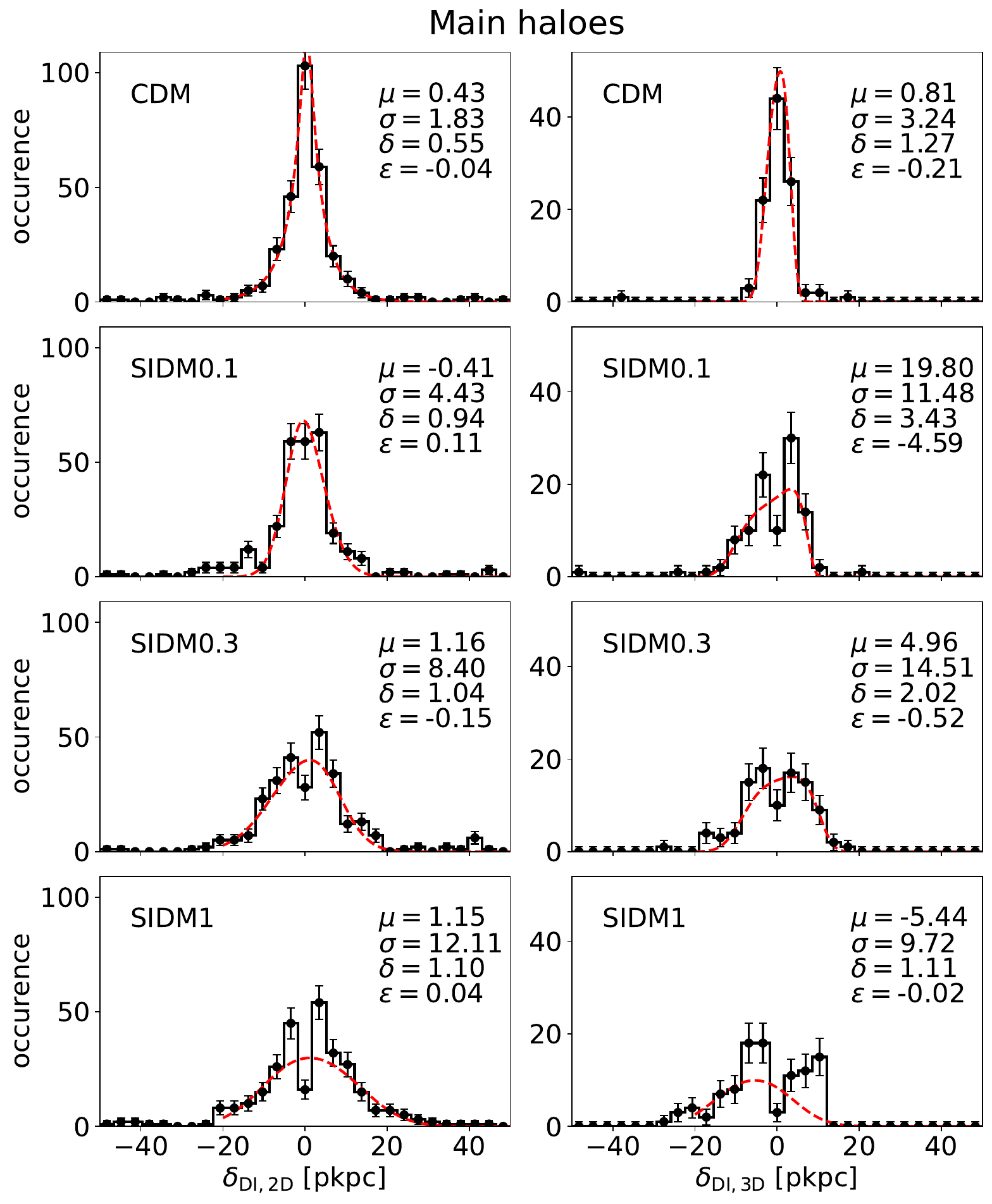}
\end{minipage}
\caption{Similar to Fig.~\ref{fig:si_distributions}, but now for the component of the vector from stellar matter to DM {\it perpendicular} to the direction of motion (see Fig.\ref{fig:beta_diagram}).}
\label{fig:di_distributions}
\end{figure*}

\begin{figure*}
\centering
\begin{minipage}{0.48\textwidth}
\includegraphics[width=\linewidth]{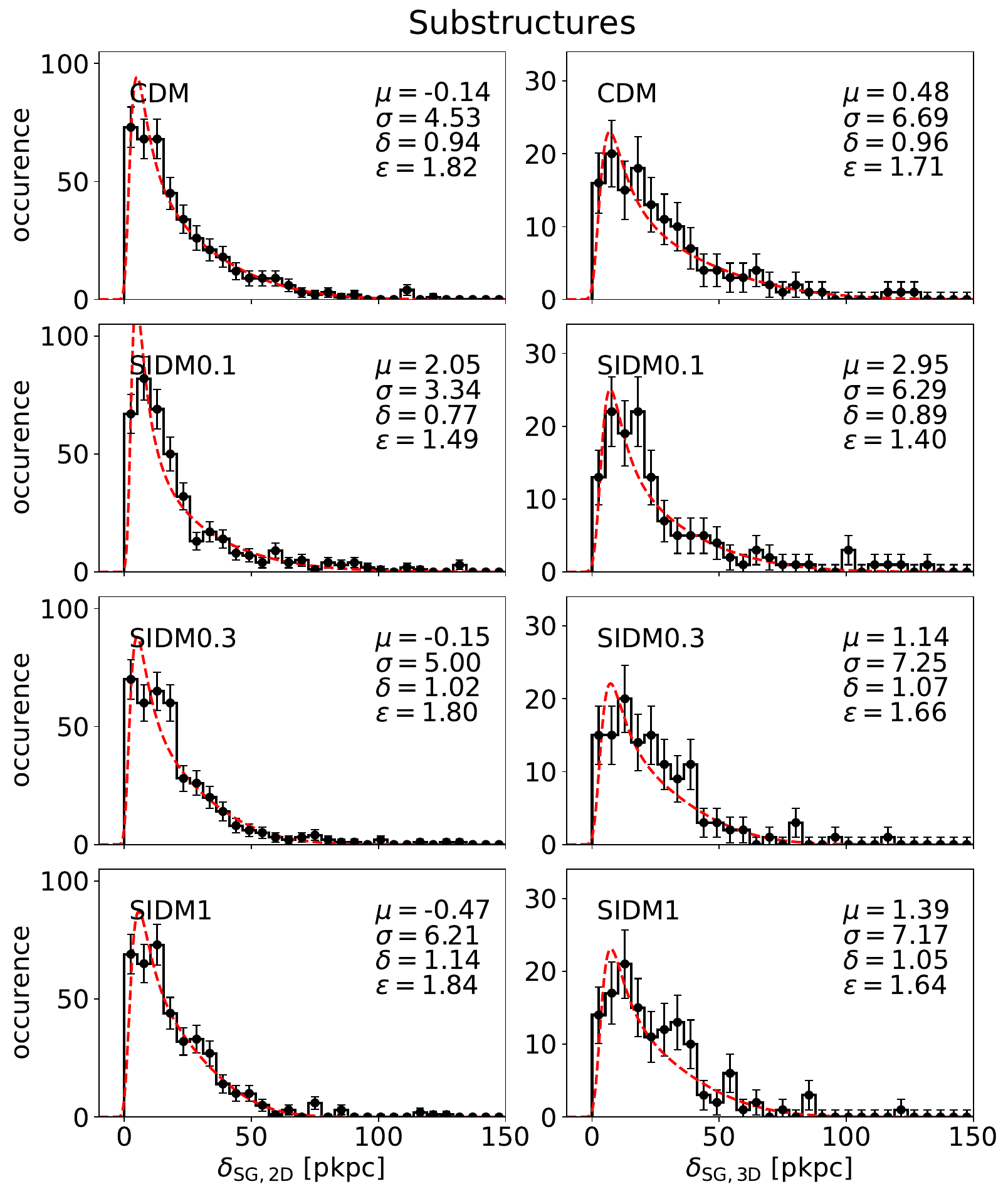}
\end{minipage}\hfill
\begin{minipage}{0.48\textwidth}
\includegraphics[width=\linewidth]{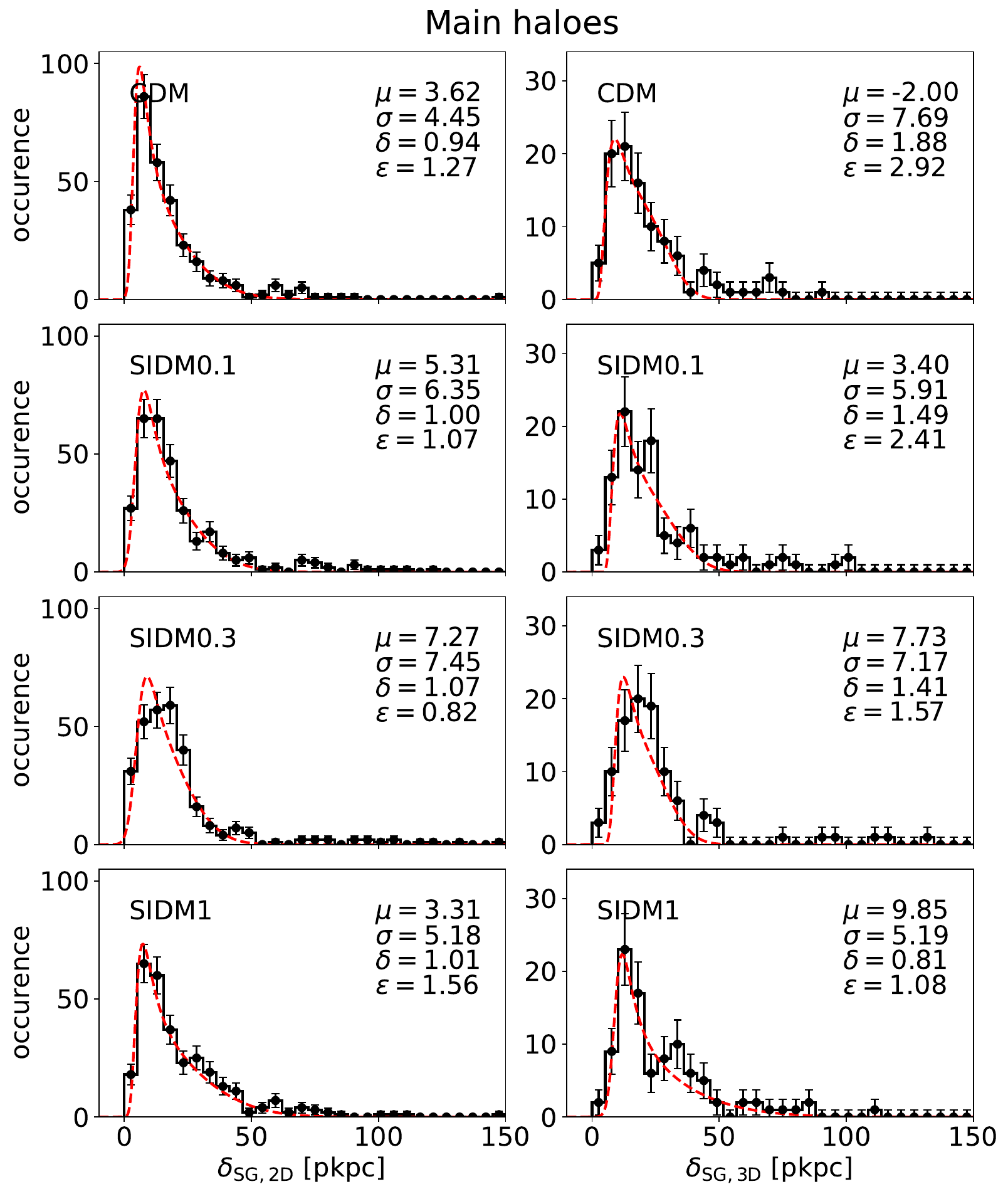}
\end{minipage}
\caption{Similar to Fig.~\ref{fig:si_distributions} and \ref{fig:di_distributions}, but now for the distance from stellar matter to the gas (see Fig.\ref{fig:beta_diagram}). Note that by definition $\dSG$ is always positive.}
\label{fig:sg_distributions}
\end{figure*}

Figures~\ref{fig:si_distributions}, \ref{fig:di_distributions} and \ref{fig:sg_distributions} show the distributions of distances used to calculate the distance ratios $\bpar$ and $\bper$ (which are themselves shown in Figures~\ref{fig:distributions_sub} and \ref{fig:beta_sim_all}). Uncertainties on the number of counts in each bin are assumed to be the square root of the counts plus one. One unexplained curiosity is that anomalously few systems have $\dDI=0$ in SIDM simulations. We cannot explain this, and merely speculate that the galaxies may be oscillating within the DM halo on a radial orbit, and thus spend very little time at pericentre.

\label{lastpage}
\bsp	
\end{document}